\documentclass[twocolumn]{aastex631}

\usepackage{amsmath}	
\usepackage{threeparttable}
\usepackage{booktabs}
\usepackage{float}

\newcommand{\msun}{{\rm M}_\odot}
\newcommand{\sbunit}{\rm erg \, s^{-1}\, cm^{-2}\,arcsec^{-2}}

\newcommand{\vect}[1]{\boldsymbol{#1}}

\newcommand{\oiii}{[\ion{O}{3}]\,$\lambda\,5008$ }

\begin{document}
\title{Resolving turbulence drivers in two luminous obscured quasars with \textit{JWST}/NIRSpec IFU}

\author[0000-0002-8739-3163]{Mandy C. Chen}
\affiliation{Cahill Center for Astronomy and Astrophysics, California Institute of Technology, Pasadena, CA 91125, USA}
\affiliation{The Observatories of the Carnegie Institution for Science, 813 Santa Barbara Street, Pasadena, CA 91101, USA}
\affiliation{Department of Astronomy and Astrophysics, The University of Chicago, Chicago, IL 60637, USA}

\author[0000-0001-8813-4182]{Hsiao-Wen Chen}
\affiliation{Department of Astronomy and Astrophysics, The University of Chicago, Chicago, IL 60637, USA}

\author[0000-0002-1690-3488]{Michael Rauch}
\affiliation{The Observatories of the Carnegie Institution for Science, 813 Santa Barbara Street, Pasadena, CA 91101, USA}

\author[0000-0002-0710-3729]{Andrey Vayner}
\affiliation{IPAC/Caltech, 1200 E. California Boulevard, Pasadena, 91125, CA, USA}

\author[0000-0003-3762-7344]{Weizhe Liu}
\affiliation{Department of Astronomy and Joint Space-Science Institute, University of Maryland, College Park, MD 20742, USA}

\author[0000-0002-1608-7564]{David S. N. Rupke}
\affiliation{Department of Physics, Rhodes College, Memphis, TN 38112, USA}
\affiliation{Zentrum für Astronomie der Universität Heidelberg, Astronomisches Rechen-Institut, Mönchhofstr 12-14, D-69120 Heidelberg, Germany}

\author[0000-0002-5612-3427]{Jenny E. Greene}
\affiliation{Department of Astrophysical Sciences, Princeton University, 4 Ivy Lane, Princeton, NJ 08544, USA}

\author[0000-0001-6100-6869]{Nadia L. Zakamska}
\affiliation{Department of Physics and Astronomy, Bloomberg Center, Johns Hopkins University, Baltimore, MD 21218, USA}
\affiliation{Institute for Advanced Study, Princeton, NJ 08540, USA}

\author[0000-0003-2212-6045]{Dominika Wylezalek}
\affiliation{Zentrum für Astronomie der Universität Heidelberg, Astronomisches Rechen-Institut, Mönchhofstr 12-14, D-69120 Heidelberg, Germany}

\author[0000-0003-2390-7927]{Guilin Liu}
\affiliation{CAS Key Laboratory for Research in Galaxies and Cosmology, Department of Astronomy, University of Science and Technology of China, Hefei 230026, People's Republic of China}
\affiliation{School of Astronomy and Space Science, University of Science and Technology of China, Hefei 230026, People's Republic of China}

\author[0000-0002-3158-6820]{Sylvain Veilleux}
\affiliation{Department of Astronomy and Joint Space-Science Institute, University of Maryland, College Park, MD 20742, USA}

\author[0000-0001-5783-6544]{Nicole P. H. Nesvadba}
\affiliation{Université de la Côte d'Azur, Observatoire de la Côte d'Azur, CNRS, Laboratoire Lagrange, Bd de l'Observatoire, CS 34229, Nice cedex 4 F-06304, France}

\author[0000-0002-6948-1485]{Caroline Bertemes}
\affiliation{Zentrum für Astronomie der Universität Heidelberg, Astronomisches Rechen-Institut, Mönchhofstr 12-14, D-69120 Heidelberg, Germany}





\begin{abstract}
In this {\it Letter}, we investigate the turbulence and energy injection in the extended nebulae surrounding two luminous obscured quasars, WISEA J100211.29$+$013706.7 ($z=1.5933$) and SDSS J165202.64$+$172852.3 ($z=2.9489$). Utilizing high-resolution data from the NIRSpec IFU onboard the \textit{James Webb Space Telescope}, we analyze the velocity fields of line-emitting gas in and around these quasars and construct the second-order velocity structure functions (VSFs) to quantify turbulent motions across different spatial scales. 
Our findings reveal a notable flattening in the VSFs from $\approx\!3$ kpc up to a scale of 10--20 kpc, suggesting that energy injection predominantly occurs at a scale $\lesssim$10 kpc, likely powered by quasar outflows and jet-driven bubbles. 
The extended spatial range of flat VSFs may also indicate the presence of multiple energy injection sources at these scales. 
For J1652, the turbulent energy in the host interstellar medium (ISM) 
is significantly higher than in tidally stripped gas,
consistent with the expectation of active galactic nucleus (AGN) activities stirring up the host ISM. Compared to the VSFs observed on spatial scales of 10--50 kpc around lower-redshift UV-bright quasars, these obscured quasars exhibit higher turbulent energies in their immediate surroundings, implying different turbulence drivers between the ISM and halo-scale gas. 
Future studies with an expanded sample are essential to elucidate further the extent and the pivotal role of AGNs in shaping the gas kinematics of host galaxies and beyond. 

\end{abstract}



\section{Introduction} 
\label{sec:intro}
Over the past two decades, mounting evidence has shown that feedback from active galactic nuclei (AGNs) is instrumental in driving a galaxy's growth or quiescence via regulating the ebbs and flows of energy and materials in and around galaxies \citep[see reviews by, e.g.,][]{Fabian2012,Heckman2014, Harrison2018}. Despite the important implications of AGN feedback on many aspects of cosmic history, direct observational evidence and a detailed understanding of the mechanisms by which energy is transferred from supermassive black holes to the galaxy- and halo-scale gas remain elusive. This knowledge gap is due in part to the observational challenge of resolving the multiscale gas physics around AGNs and mapping the kinematic and energetic properties over a wide range of spatial scales. Advances in integral field unit (IFU) spectroscopy, particularly with wide-field, high-throughput instruments like the Multi-Unit Spectroscopic Explorer \citep[MUSE;][]{Bacon2010} on the Very Large Telescope (VLT) and the Keck Cosmic Web Imager \citep[KCWI;][]{Morrissey2018} on the Keck Telescopes, have opened new avenues for resolving a variety of dynamical processes in unprecedented detail \citep[e.g.][]{Johnson2018,Rupke2019,Chen2019,Johnson2022}. Moreover, the IFU mode of the Near-Infrared Spectrograph \citep[NIRSpec;][]{Jakobsen2022} onboard the \textit{James Webb Space Telescope} \citep[\textit{JWST};][]{Gardner2006} has recently opened a new window into the gaseous environments at high redshifts by offering spatially-resolved properties of rest-frame optical nebular lines. 

This {\it Letter} presents an investigation into the kinetic properties of the extended nebulae surrounding two obscured quasars, WISEA J100211.29+013706.7 and SDSS J165202.64+172852.3 (hereafter XID2028 and J1652, respectively). The study utilizes observations obtained with \textit{JWST}/NIRSpec IFU. Both quasars are part of the Early Release Science program Q3D \citep[PI: D. Wylezalek;][]{Wylezalek2022}, and provide exceptional laboratories for studying AGN feedback due to their high luminosities and the presence of intense black hole activities. We focus particularly on the second-order velocity structure function (VSF), $S_2$, which measures the scale-dependent velocity variance following
\begin{equation}
    S_2(r)=\langle \vert \vect{v}(\vect{x}) - \vect{v}(\vect{x}+\vect{r}) \vert^2 \rangle. 
	\label{eq:vsf}
\end{equation}
Here, $\vect{x}$ represents a location in the velocity map, and $\vect{r}$ is the vector denoting the displacement between two points separated by $\vect{r}$ \citep[e.g.,][]{Frisch1995}. The angle brackets denote an average over all two-point pairs separated by a given $\vect{r}$.  With observational data, both velocity and distance measurements are limited to projected quantities.  Therefore, the VSFs presented in this {\it Letter} are determined based on line-of-sight (LOS) velocities and the projected separation $r_{\rm proj}$ in the plane of the sky as discussed further in Section \ref{sec:vsf_measurement} below. VSFs provide a key diagnostic for understanding the driving, cascade, and dissipation of turbulent kinetic energy. For example, in the case of Kolmogorov turbulence, which applies to homogeneous, isotropic, and incompressible flows, energy cascades from large to small scales, producing a characteristic power-law slope of 2/3 in $S_2$ within the inertial range \citep{Kolmogorov1941}. VSFs have been extensively used to investigate the dynamical conditions of the interstellar medium (ISM) \citep[e.g., ][]{Wen1993, Ossenkopf2002,Federrath2013,Arthur2016,Padoan2016,Melnick2021,Garcia-Vazquez2023}, the circumgalactic medium 
(CGM; \citealt{Rauch2001,Chen2023,Chen2024}), as well as the intracluster medium (ICM) in nearby cool-core clusters \citep[][]{Li2020,Ganguly2023}, offering insights into the turbulent energy cascade in a wide range of gaseous environments. 

Here we present VSF measurements of line-emitting nebulae around XID2028 and J1652 using spatially resolved velocity maps from \textit{JWST}/NIRSpec IFU observations.  With a bolometric luminosity of $\approx 10^{46.3}$ erg s$^{-1}$ and $\approx 10^{47.7}$ erg s$^{-1}$, respectively, XID2028 and J1652 are among the brightest quasars known and are hosted by massive galaxies with a stellar mass $\gtrsim 10^{11}\msun$ \citep{Perna2015,Brusa2018,Zakamska2019}. 
Located at the epoch of peak black hole accretion, with XID2028 at $z\approx 1.6$ and J1652 at $z\approx 3$, these quasars represent prime candidates for studying efficient AGN feedback. By examining the VSFs in the outflows, in the systemic gas, and in tidally stripped gas of these quasar hosts, we aim to characterize the scale of turbulence and to gain a better understanding of the underlying mechanisms driving these chaotic motions. Additionally, we contextualize our findings by comparing the results with similar measurements obtained in lower-redshift quasars.

Throughout this {\it Letter}, we adopt a standard $\Lambda$CDM cosmology with $H_0$ = 70 km s$^{-1}$ Mpc$^{-1}$, $\Omega_{\rm M}$=0.7, and $\Omega_{\rm vac}$=0.3 when deriving distances and luminosities.  All distances presented here are physical/proper distances unless otherwise specified. The corresponding projected distances per arcsecond at the redshifts of XID2028 ($z=1.5933$) and J1652 ($z=2.9489$) are 8.471 kpc and 7.741 kpc, respectively.  


\section{Observations and data analysis}
\label{sec:data}
Both XID2028 and J1652 were observed by \textit{JWST}/NIRSpec IFU with a nine-point dither pattern to improve the spatial sampling. Utilizing the NIRSpec IFU data, we constructed a spatially-resolved LOS velocity map, employing the \oiii emission line as a tracer of gas motions. This section provides a brief description of the observation setup and the procedures for velocity line fitting.

\subsection{\textit{JWST}/NIRSpec IFU observations}
\label{sec:JWST_data}
XID2028 was observed in November 2022 and the [\ion{O}{3}] emission was targeted with the G140H/F100LP grating/filter pair with a total exposure time of 2.95 hours. For J1652, the observations were taken in July 2022 with 4.6 hrs of total on-target exposure and the G235H/F170LP configuration to cover the [\ion{O}{3}] line \citep{Wylezalek2022, Vayner2023, Vayner2024}. 

Data reduction for both fields was performed using the standard \textit{JWST} pipeline to process dark current subtraction, bias subtraction, cosmic-ray removal, background subtraction, and flux calibration. A custom routine was employed to project the data cubes onto a $0\farcs05$ spatial grid, and the final spectral sampling is 0.0235 $\micron$ per spectral pixel for XID2028 and 0.0396 $\micron$ for J1652 \citep{Vayner2023, Veilleux2023, Vayner2024}. 
Due to a significant amount of artifacts in the spectrum and the spatial undersampling, the effective spatial resolution of the data cubes is characterized by a full-width-at-half-maximum (FWHM) of $\approx 0\farcs2$, corresponding to a physical scale of $\approx 1.5$--1.7 kpc at the redshifts of XID2028 and J1652. 

\begin{figure*}
\centering
    \includegraphics[width=0.85\textwidth]{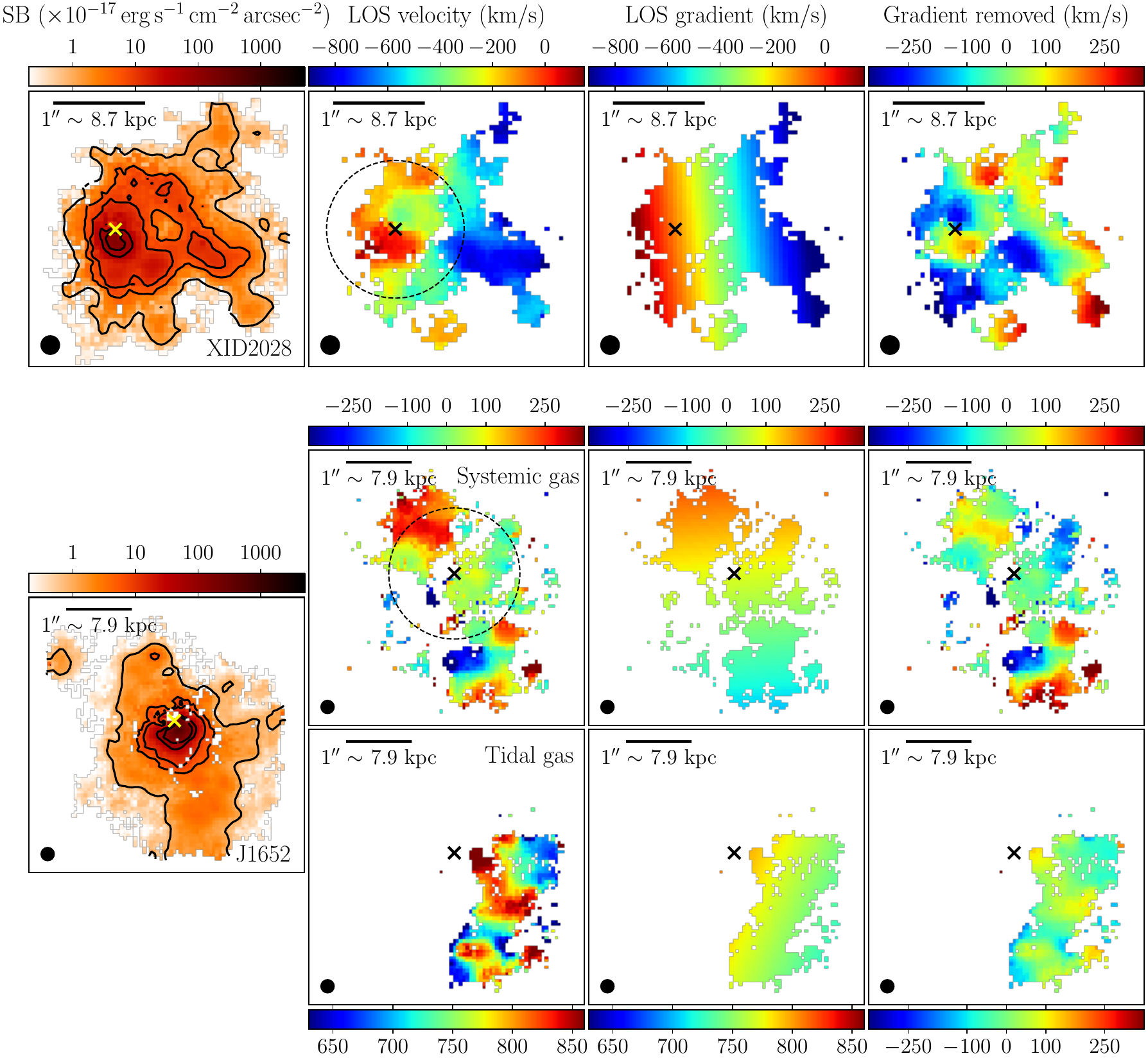}
    \caption{{\it From left to right:} the columns display the continuum- and quasar-subtracted [\ion{O}{3}] narrow-band image, the LOS velocity map, the best-fit unidirectional velocity gradient, and the residual LOS velocity map with the unidirectional gradient removed. The contours in the first column correspond to surface brightness levels of [5, 10, 50, 100]$\times 10^{-17}\sbunit$. {\it The top row} presents the field of XID2028 at $z=1.5933$, while {\it the middle and bottom rows} show the signal from the systemic and tidally stripped gas, respectively, in the field of J1652 at $z=2.9489$. Note that the velocity panels only highlight spaxels included in the VSF measurements (see Section \ref{sec:vsf_measurement}), and the narrow-band image of J1652 shows the combined flux from outflows, systemic gas, and tidally stripped gas (see Section \ref{sec:velocity_fitting}). The cross in each panel indicates the location of the quasar, and the black circle represents the size of the PSF in the data. The dashed circles in the second column represent the division of the inner versus outer regions for the VSF measurements (see Section \ref{sec:result_vsf_inner_vs_outer}). Both fields show extended emission reaching up to $\sim 20$ kpc and the associated velocity fields appear to be turbulent. 
    }
    \label{fig:nb_and_velocity}
\end{figure*}

We use \texttt{q3dfit}\footnote{https://q3dfit.readthedocs.io/stable/} (v1.0.0) to model and remove the quasar emission \citep{Rupke2023_q3d}.  A low-order polynomial continuum attributed to the quasar host is also modeled and removed, resulting in a continuum- and quasar-subtracted emission-line-only cube for both fields \citep[see][for details]{Vayner2023}. In the first column of Figure \ref{fig:nb_and_velocity}, we show the continuum- and quasar-subtracted narrow-band images focusing on the emission from \oiii, which was optimally extracted based on a 3D mask in the position-velocity space \citep[see e.g.][]{Borisova2016, Chen2023}. These narrow-band images are independent of the subsequent model-fitting procedure, and the map of J1652 displays the combined flux contributions from outflows, systemic gas, and tidally stripped gas. Flux maps of individual gas components based on model fitting are shown in Figure 1 of \cite{Vayner2023}. 

\subsection{Velocity measurements}
\label{sec:velocity_fitting}
To obtain the spatially-resolved LOS velocity based on the [\ion{O}{3}] line, we follow the procedure detailed in \cite{Chen2023} to conduct the line fitting using MCMC including up to three independent Gaussian components. For the purpose of this work, we only fit the velocity field based on the [\ion{O}{3}]$\lambda\lambda$4960,5008 lines, which are redshifted to an observed wavelength of $\approx1.3~\micron$ for XID2028 and $\approx2~\micron$ for J1652. We include both [\ion{O}{3}] lines to enhance the constraining power of the data and adopt a fixed line ratio of [\ion{O}{3}]$\lambda4960$/[\ion{O}{3}]$\lambda5008$ = 1/3 during fitting. We present the LOS velocities with respect to the systemic quasar redshift of each field, which was measured based on the narrow [\ion{O}{3}] emission from the central region as described in \cite{Wylezalek2022} and \cite{Veilleux2023}.

The fitting results of XID2028, in line with \cite{Veilleux2023}, show that the spaxels can be adequately described using one or two Gaussian components with the majority ($\approx 85$\%) of the spaxels requiring only one component. After correcting for instrument line broadening, almost all spaxels (whether modeled with one or two components) exhibit a relatively narrow line width of $\sigma_v<300$ km/s, with only $<3\%$ of spaxels requiring a Gaussian component with $\sigma_v>300$ km/s.  Some of these broad components can be attributed to residuals of imperfect continuum subtraction and/or fast outflows in localized regions, and these can contribute to elevated velocity uncertainties.  Therefore, we exclude these broad components in subsequent analyses. We adopt a flux-weighted mean velocity centroid for spaxels that have two Gaussian components, and present the LOS velocity map of each quasar nebula in Figure \ref{fig:nb_and_velocity}, highlighting only the spaxels included in the VSF measurement (see Section \ref{sec:vsf_measurement} below). As evident from the velocity map, the nebula surrounding XID2028 exhibits a predominantly blue-shifted velocity field with a median LOS velocity across all spaxels of $\approx -400$ km/s from the quasar redshift. Previous multiwavelength observations of this quasar have identified XID2028 as a host of a massive, multiphase outflow \citep[e.g.][]{Brusa2015,Perna2015,Brusa2018} where the ionized warm phase is traced by \oiii emission. 

In contrast, the [\ion{O}{3}] line profiles of J1652 across the NIRSpec IFU field of view are more complex. The majority of the spaxels in the field of J1652 require three components for an accurate fit, in agreement with the independent fitting results presented in \cite{Vayner2023}.  As detailed in \cite{Vayner2023}, these components can be grouped into three distinct physical origins: galactic outflows characterized by a Gaussian line width $\sigma_v>300$ km/s, systemic quasar host gas with a velocity $|\Delta v|<500$ km/s from the quasar redshift and a line width $\sigma_v<300$ km/s, and tidally stripped gas from companion galaxies, which exhibits $|\Delta v|>500$ km/s relative to the quasar redshift and a line width $\sigma_v<300$ km/s. Following the same classification criteria, we construct velocity maps for these three components, which are consistent with results presented in \cite{Vayner2023}. In Figure \ref{fig:nb_and_velocity}, we show the LOS velocity maps of the systemic gas and the tidally stripped gas.  Due to its spatial confinement, we do not include the outflow component in subsequent VSF measurements. 

For the purpose of this work, we only focus on VSF measurements derived from velocity centroids across nebulae.  While line widths offer supplementary insights into gas motions, their interpretation as indicators of turbulence is complicated by LOS bulk flows and PSF smoothing.  Future work, incorporating careful comparisons with numerical simulations, will further elucidate potential synergies between the constraints provided by line widths and VSFs. 

\subsection{Constructing a second-order VSF}
\label{sec:vsf_measurement}
To obtain robust VSF measurements, we exclude spaxels with a velocity centroid uncertainty $>45$ km/s. In addition, we also model and remove a unidirectional velocity gradient to mitigate the influence of coherent bulk flows on the VSFs, as described in Section 4.2 of \cite{Chen2023}. 

As discussed above, the nebula in XID2028 is known to be associated with a large-scale outflow, which is evidenced by the predominantly blue-shifted velocity field. Indeed, the best-fit gradient for XID2028 is substantial, $\approx 65$ km/s/kpc (see Figure \ref{fig:nb_and_velocity}), likely tracing the bulk motion of the outflow region in the warm ionized gas. This simple unidirectional velocity gradient effectively removes the dominant contribution of coherent bulk motion to the apparent VSF.  While in future work we will explore using more physically motivated models to better account for bulk flows, a more sophisticated approach is unlikely to significantly impact the VSFs.  This is supported by results from J1652 (see below) and systems studied by \citep{Chen2024}, where gradients with small amplitudes have minimal effect, yielding consistent VSF measurements before and after gradient removal.

For J1652, our analysis centered on the components of the systemic gas and tidally stripped gas.  The outflow component with a broad line width is confined to the innermost region ($\lesssim5$ kpc from the quasar) and lacks sufficient spatial coverage for a robust VSF measurement. In the middle and bottom rows of Figure \ref{fig:nb_and_velocity}, we show the LOS velocity, their associated best-fit unidirectional gradient, and the residual maps with the gradient removed for spaxels that are included in the VSF measurements. Unlike XID2028, the systemic and the tidal components of J1652 exhibit only a modest velocity gradient of $\approx 13$ km/s/kpc and $\approx 5$ km/s/kpc, respectively. The tidal component consists of three distinct nebulae that are disjoint in the position-velocity space \citep[see the bottom row of Figure 1 in][]{Vayner2023}. 
For the VSF measurements, we focus on the most extended, southwest nebula.

For both fields, we measure the second-order VSF as a function of projected distance separation, $r_{\rm proj}$, in the plane of the sky. We designate the observed VSF as $S_2^\prime$ to include the ``seeing" smoothing effect in the data.  Generally, the spatial correlation due to the PSF smoothing preferentially removes power from small scales, causing the observed VSF, $S_2^\prime$, to appear steeper than the intrinsic one, $S_2$.  This effect is the most pronounced at scales of $r_{\rm proj}\lesssim2\times$FWHM and can be explicitly incorporated into the model $S_2^\prime$ as a convolution between the PSF smoothing kernel and the intrinsic, unsmoothed velocity field \citep[see Equations 2--7 in][]{Chen2023}. To robustly estimate the uncertainty in the measured $S_2^\prime$, considering both the velocity measurement uncertainties and the correlation among spatially adjacent pixels, we follow the modified bootstrap method with 1000 samples as outlined in \cite{Chen2023}. Additionally, we calculated $S_2^\prime$ using both the directly measured velocity maps and the residual maps after removing the unidirectional gradient to assess the potential contribution of bulk flow.

\section{Results}
In this section, we present the VSF measurements of XID2028 and J1652.  We discuss the general trends in gas turbulence for both fields as elucidated by the properties of the VSFs, and contextualize the results of XID2028 and J1652 with previous studies of low-redshift quasar nebulae on larger spatial scales using VLT/MUSE data. 

\subsection{VSFs in line-emitting gas around XID2028 and J1652}
\begin{figure}
    \includegraphics[width=\linewidth]{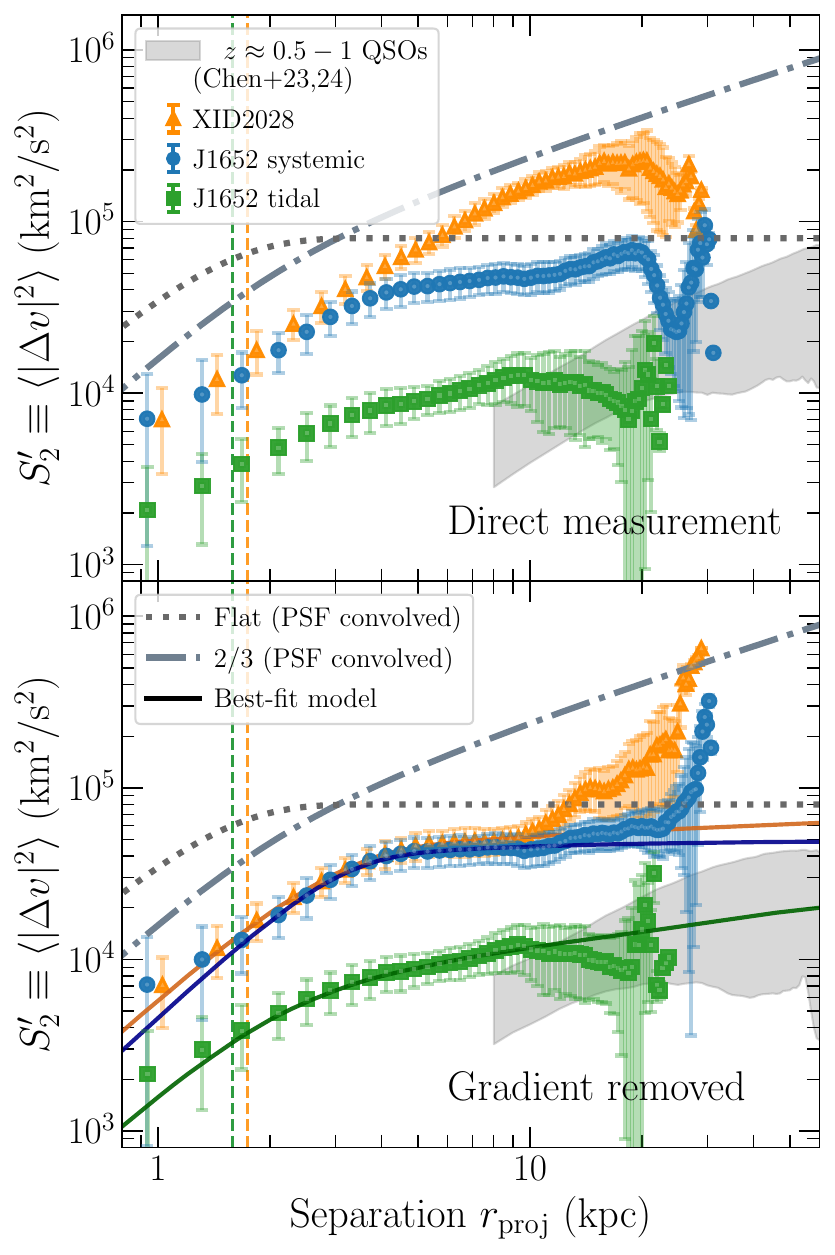}
    \caption{The second-order VSF, $S_2^\prime(r)$. The top panel presents measurements constructed from the directly measured line-of-sight velocity maps of XID2028 and J1652, whereas the bottom panel shows the results with the velocity gradient subtracted maps (see columns 2--4 of Figure \ref{fig:nb_and_velocity}). Vertical dashed lines indicate the FWHM of the PSF for the two fields.  The gray dot-dashed curve represents $S_2^\prime(r)$ with a Kolmogorov slope of 2/3, convolved with a mean PSF of the two fields.  Gray-shaded regions represent previous measurements for a sample of eight extended nebulae around UV-luminous quasars at $z\approx 0.5$--1 \citep{Chen2023,Chen2024}. Within the overlapping scales of approximately 10--20 kpc, the measurements from XID2028 and the systemic gas of J1652 display a higher amplitude in contrast to the UV quasar sample. Conversely, the VSF of the tidally stripped gas in J1652 is $\sim 10$ times lower in amplitude, in good agreement with the lower-redshift UV sample.
    }
    \label{fig:s2}
\end{figure}

Figure \ref{fig:s2} shows the VSF measurements for XID2028 and J1652 across projected distance scales from $\approx 1$ kpc to $\approx 30$ kpc. 
At large scales ($\gtrsim20$ kpc), measurement uncertainties increase as a result of limited two-point sampling and more uncertain velocity measurements in the outskirts of these nebulae \citep[see e.g.,][for detailed discussions on the impact of limited nebula size]{Garcia-Vazquez2023,Chen2024}. Consequently, we refrain from interpreting measurements beyond 20 kpc. For XID2028, the substantial velocity gradient of $\approx 65$ km/s/kpc, likely tracing the bulk motion in the outflows, contributes significantly to the power at $\gtrsim4$ kpc.  After removing this gradient, the observed $S_2^\prime$ of XID2028 exhibits an overall flat slope from $r_{\rm proj}\approx 4$ kpc up to 10 kpc.  Conversely, for J1652, the VSFs for both the systemic and tidal components are consistent before and after removing the velocity gradient, reflecting the minor impact of the gradient due to its small amplitude in both components. We also test the VSF of XID2028 with excluding a 3 kpc radius inner area around the quasar, where a potential rotating disk is detected \citep[][]{Cresci2023,Veilleux2023}.  Results are consistent with and without this central region, and we opt to include the central region for simplicity.

After removing the unidirectional velocity gradient, the overall profile of the observed second-order VSFs can be characterized by a steep slope at smaller scales ($r_{\rm proj}\lesssim 2$--4 kpc) and a notable flattening beyond $\approx 4$ kpc. To quantify these observed trends in the VSFs, we use a double power law model, parameterized as: 
\begin{equation}
\label{eq:broken_powerlaw}
    S_2 = 
    \begin{cases} 
    a\,r_{\rm proj}^{\gamma_1} & \text{if } r_{\rm proj} \leq r_{\rm b} \\
    (a\,r_{\rm b}^{\gamma_1}/r_{\rm b}^{\gamma_2})\,r_{\rm proj}^{\gamma_2} & \text{if } r_{\rm proj} > r_{\rm b}
    \end{cases}
\end{equation}
where $a$, $r_{\rm b}$, $\gamma_1$, and $\gamma_2$ are the model parameters representing the normalization, the break radius, and the slopes of the power law below and above the break radius, respectively. 
The model is convolved with the appropriate PSF for each field following Equations 2--7 of \cite{Chen2023} to derive $S_2^{\prime}$. The fitting was conducted over the range from the PSF FWHM of each field (vertical dashed lines in Figure \ref{fig:s2}) to 10 kpc.  Table \ref{tab:epsilon_summary} lists the median values along with the 16$^{\rm th}$ and 84$^{\rm th}$ percentile uncertainties obtained by fitting the 1000 modified bootstrap VSF samples (see Section \ref{sec:vsf_measurement} above).  The best-fit median models are shown by solid curves in the bottom panel of Figure \ref{fig:s2}. 

\begin{deluxetable*}{lccccccccc}
         \tablenum{1}
        \tablecaption{Summary of the VSF double power law fitting (see Equation \ref{eq:broken_powerlaw}) and the measurements for the energy cascade rate per unit mass (see Equation \ref{eq:epsilon}) in different systems at different physical scales; results are shown as median values with the 16$^{\rm th}$ and 84$^{\rm th}$ percentile uncertainties.}  
	\label{tab:epsilon_summary}
        \tablewidth{0pt}
        \tablehead{
        \colhead{Field name} & \colhead{$r_{\rm b}$} & \colhead{$\gamma_1$} & \colhead{$\gamma_2$} & \colhead{$\epsilon_{\rm 4kpc}$} & \colhead{$\epsilon_{\rm 10kpc}$} & \colhead{$\epsilon_{\rm 50kpc}$} & \colhead{$\langle\epsilon\rangle_{\rm inner}^a$} & \colhead{$\langle\epsilon\rangle_{\rm outer}$} & \colhead{$\langle\epsilon\rangle_{\rm outer}$/$\langle\epsilon\rangle_{\rm inner}$}\\
        \cmidrule(lr){2-2}\cmidrule(lr){5-10}
         & (kpc) &  & \multicolumn{7}{c}{(cm$^2$ s$^{-3}$)}}
        \startdata
		MUSE quasars$^b$ & -- & -- & -- & -- & 0.04$^{+0.04}_{-0.02}$ & 0.03$^{+0.07}_{-0.02}$ & -- & -- & -- \\ 
		XID2028$^c$ & 3.5$^{+1.6}_{-1.3}$ & 1.2$^{+2.8}_{-0.8}$ & 0.07$^{+0.21}_{-0.07}$ & 1.2$^{+0.3}_{-0.3}$ & 0.8$^{+0.2}_{-0.2}$ & -- &0.7$^{+0.2}_{-0.2}$ &1.2$^{+0.4}_{-0.4}$ &1.6$^{+1.0}_{-0.6}$ \\ 
		J1652 systemic$^d$ & 3.1$^{+1.3}_{-0.6}$ & 2.3$^{+2.7}_{-1.6}$ & 0.0$^{+0.2}_{-0.0}$ & 1.3$^{+0.5}_{-0.4}$ & 0.6$^{+0.2}_{-0.2}$ & -- &0.5$^{+0.2}_{-0.2}$ & 1.8$^{+0.6}_{-0.4}$ &3.6$^{+2.6}_{-1.5}$ \\ 
		J1652 tidal & 2.1$^{+3.5}_{-1.2}$ & 2.8$^{+2.2}_{-2.7}$ & 0.2$^{+0.2}_{-0.2}$& 0.11$^{+0.05}_{-0.04}$ & 0.07$^{+0.04}_{-0.03}$ & -- &-- & -- &-- \\ 
        \enddata 
        \tablecomments{\\
        $^a$ The angle bracket denotes the median $\epsilon$ value taken between the scales of $4$ kpc and 10 kpc.\\
        $^b$ A sample of eight extended nebulae surrounding seven luminous quasars at $z\approx 0.5$--1 with bolometric luminosities of $L_{\rm bol}\gtrsim10^{47}$ erg s$^{-1}$, detected by VLT/MUSE \citep{Chen2023, Chen2024}. \\
        $^c$ For XID2028, the inner region is defined as the circular area with a radius of $\lesssim6.5$ kpc from the quasar location (see text and Figure \ref{fig:nb_and_velocity}).\\
        $^d$ For J1652 systemic gas, the inner region is defined as the circular area with a radius of $\lesssim8$ kpc from the quasar location (see text and Figure \ref{fig:nb_and_velocity}).}
\end{deluxetable*}

For the XID2028 outflows, J1652 systemic gas, and J1652 tidally stripped gas, the best-fit break radii range from approximately 2 to 4 kpc. Above these break radii, the power law slope $\gamma_2$ approaches zero, indicating a flattening of the VSFs. Below the break radii, the slope $\gamma_1$ is steeper, although the estimates are associated with large uncertainties due to the limited fitting range.



Flattening in the VSF slopes on large scales is a strong indicator of energy injection. With the absence of energy cascades from large to small scales, a VSF flattens at $\approx 1/2$ of 
the injection scale \citep{ZuHone2016,Federrath2021}. The observed flattening of $S_2^\prime$ at $\approx 2$--4 kpc in outflows around XID2028 and the systemic and tidal gas components around J1652, therefore, suggests an energy injection scale of $\lesssim 10$ kpc in these nebulae. In addition,  
this observation aligns well with the observation that the jet-driven bubble seen in XID2028 has a size of $\sim 5$ kpc \citep{Cresci2023} and the fast outflow component in J1652 is confined to the central region with a distance $\lesssim 5$ kpc from the quasar \citep{Vayner2023,Vayner2024}, both of which are likely dominant drivers of turbulence in the host ISM and can couple the kinetic energy to the surrounding gas on scales of a few kpc. However, with the current data, we do not detect the cascade of kinetic energy to smaller scales.  The steepening observed at scales below $\approx$2–4 kpc may result from PSF smoothing effects, making it difficult to distinguish this from the effects of energy cascade. This degeneracy is reflected in the large uncertainties in the power law slope $\gamma_1$ derived from the double power law fit (see Table \ref{tab:epsilon_summary}).

Separately, a continued flatness in VSF from 10 to 20 kpc for the J1652 systemic gas implies that additional mechanisms may maintain the elevated level of kinetic energy at these scales.  
Numerical simulations show that the superposition of kinetic power spectra with energy injection at various characteristic sizes can manifest as a flat VSF \citep{Yoo2014, ZuHone2016}. This scenario is qualitatively consistent with an enhanced power observed in the outer nebula of J1652, where outflowing gas is likely to collide with tidally stripped gas from satellites and produce a highly turbulent medium (see the discussion in \S\ \ref{sec:result_vsf_inner_vs_outer}).  
However, a caveat with this scenario is that it requires a fine-tuned balance of energy injection rates across various scales to account for the observed flatness in the VSF of the J1652 systemic gas. 
Future work, particularly with numerical simulations that can resolve a diverse range of dynamical processes, will provide key lights into the nature of a flat VSF in quasar host environments.


\subsection{Turbulent energy transfer rate, $\epsilon$}

Figure \ref{fig:s2} also shows that the VSFs of XID2028 and J1652 systemic gas are approximately an order of magnitude higher in amplitude than the VSF of the J1652 tidal component. Furthermore, the VSF amplitude of the J1652 tidal gas agrees with the measurements from quasar nebulae at $z\approx0.5$--1 \citep{Chen2023, Chen2024}, which are shown by the gray shaded regions in Figure \ref{fig:s2}. 
To quantify the differences in VSF amplitudes, we calculate the mean energy transfer rate per unit mass, $\epsilon$, following the ``four-fifths law"\citep{Kolmogorov1941,Frisch1995}:
\begin{equation}
\label{eq:epsilon}
    \epsilon = \frac{5}{4}\left[\frac{|\langle \Delta v (r)^3\rangle|}{r}\right] \approx \frac{5}{4}\left[\frac{\langle |\Delta v (r)|^3\rangle}{r}\right].
\end{equation} 
Note that when the energy injection scale is smaller than 10 kpc in these nebulae as discussed above, there is a lack of kinetic energy coupling contained across larger scales. Consequently, the value of $\epsilon$ does not accurately represent the actual rate of energy transfer. On the other hand, if the observed flatness in the VSF of J1652 systemic gas between 10 and 20 kpc is a direct consequence of energy injection at multiple scales, $\epsilon$ at this scale range not only reflects the kinetic energy directly injected at a given scale but also includes energy transferred from larger scales. Keeping these caveats regarding the physical interpretation of $\epsilon$ in mind, we utilize $\epsilon$ as a measure of the level of velocity fluctuations across various scales to facilitate a quantitative discussion. 

For Kolmogorov turbulence, $\epsilon$ is a constant across scales within the inertial range.  However, as the VSFs of XID2028 and J1652 are flatter than the Kolmogorov slope, $\epsilon$ is larger at small spatial scales.  We therefore estimate $\epsilon$ at 4 kpc and 10 kpc for XID2028 and J1652 to reflect the range of $\epsilon$ in these nebulae, as listed in Table \ref{tab:epsilon_summary}. Both XID2028 outflows and J1652 systemic gas exhibit an $\epsilon$ value of $\approx 0.6$--$1.3$ cm$^2$ s$^{-3}$, which is approximately an order of magnitude larger than $\epsilon$ estimated for the J1652 tidal gas.

In addition, we contextualize the turbulent energy in these high-redshift quasar nebulae by comparing them with data from a lower-redshift sample of quasar nebulae at $z\approx0.5$--1 \citep{Chen2023, Chen2024}, as introduced above. This sample comprises seven quasars, whose bolometric luminosities are comparable to those of XID2028 and J1652, as noted in Table \ref{tab:epsilon_summary}. Unlike the obscured quasars, the quasars in the lower-redshift sample are UV-bright, indicating minimal dust extinction near the active supermassive black hole. This characteristic suggests that UV-bright quasars may represent a later evolutionary stage in the radiative phase of a quasar’s lifecycle. In contrast, obscured quasars are typically linked with the initial ``blowout phase" of quasar evolution \citep[e.g.][]{Wylezalek2022}.

At 10 kpc, which is a scale accessible via both the \textit{JWST}/NIRSpec IFU and VLT/MUSE data, the estimated $\epsilon$ for XID2028 outflow and J1652 systemic gas exceeds that of the $z\approx0.5$--1 quasar nebulae by more than tenfold. On the other hand, the estimated $\epsilon$ for J1652  tidal component is 0.07$^{+0.04}_{-0.03}$ cm$^2$ s$^{-3}$ at the same scale, aligning closely with the $\epsilon=0.04^{+0.04}_{-0.02}$ cm$^2$ s$^{-3}$ for the lower-redshift nebulae. This agreement suggests that the tidal gas in J1652 may originate from regions further from the quasar center, similar to the ionized gas on 10s of kpc scales in the lower-redshift sample. As we will discuss further in Section \ref{sec:discussion_conclusion} below, the markedly higher $\epsilon$ in XID2028 outflows and J1652 systemic gas highlights the boosted kinetic energy in these more spatially-confined nebulae.

\subsection{VSFs in the inner and outer regions of the nebulae}
\label{sec:result_vsf_inner_vs_outer}
\begin{figure}
    \includegraphics[width=\linewidth]{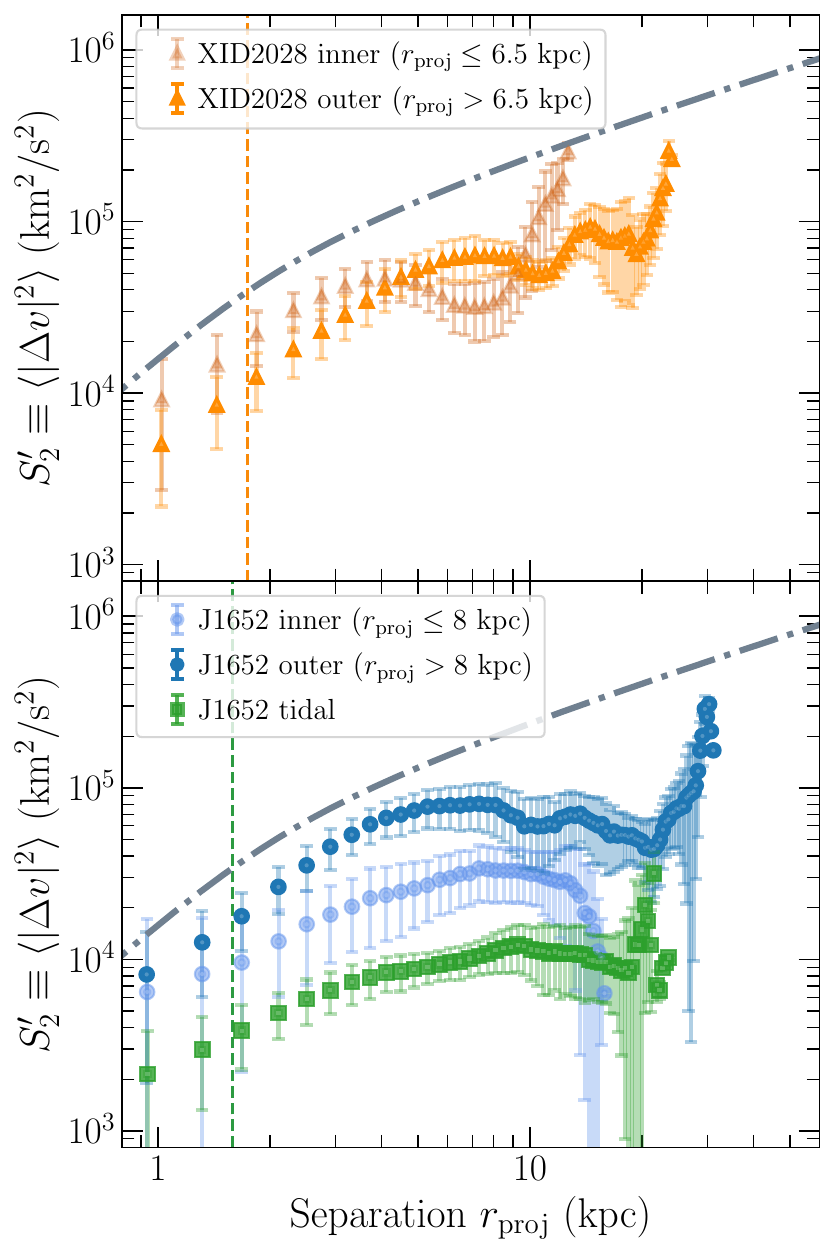}
    \caption{The second-order VSF, $S_2^\prime(r)$, measured for the inner and outer regions of the extended nebulae (see dashed circles in the second column of Figure \ref{fig:nb_and_velocity}).  {\it The top (bottom) panel} shows the results from the XID2028 (J1652) field.
    We also include the $S_2^\prime(r)$ of J1652's tidal gas for comparison. Vertical dashed lines represent the FWHM of the PSF for the two fields. The gray dot-dashed curve illustrates $S_2^\prime(r)$ modeled with a Kolmogorov 2/3 slope, after convolution with the average PSF size of the two fields. The VSFs measured from the inner and outer regions of the XID2028 field display comparable amplitudes, while the inner VSF of the J1652 systemic gas exhibits a lower amplitude compared to its outer region, suggesting more energetic turbulent motions in the outer regions of J1652. Additionally, both the inner and outer regions of the J1652 systemic gas show a higher VSF amplitude compared to the tidal component. 
    }
    \label{fig:s2_inner_vs_outer}
\end{figure}

To further dissect possible energy injection sources in XID2028 and J1652 systemic gas, we divide the nebulae into inner and outer regions. Specifically, for the XID2028 outflows and J1652 systemic gas, the inner region is a circular area with radii less than 6.5 kpc and 8 kpc, respectively, from the quasar center, as shown by the black dashed circles in Figure \ref{fig:nb_and_velocity}. The outer region consists of spaxels exterior to these circles, with the division chosen to roughly equalize the spaxel count between the regions. We conduct separate VSF measurements for inner and outer regions, as presented in Figure \ref{fig:s2_inner_vs_outer}. 

Notably, XID2028 outflows and the systemic component of J1652 exhibit different trends.  For XID2028,  the VSFs of the inner and outer regions are consistent within measurement uncertainties, albeit additional substructures (e.g., at $\approx 7$ kpc) likely due to variance within different subregions of this off-centered nebula. In contrast, the systemic gas of J1652 yields a consistently higher amplitude in the VSF measurement in the outer region than in the inner region. The median $\epsilon$ value between scales of $4$ kpc and $10$ kpc in the outer region is $\approx 3.6\times$ higher than that in the inner region (see Table \ref{tab:epsilon_summary}), indicating a higher level of turbulent motions in the outer region of the systemic gas of J1652.  This trend is unexpected if we assume that the main source for energy injection is the central quasar, and that turbulence is more intense closer to the quasar and diminishes further out, similar to the observed trend in nearby cool-core clusters beyond 10s of kpc scales from the central black hole \citep[][]{Zhuravleva2014}.

The measurements shown in Figure \ref{fig:s2_inner_vs_outer} suggest an additional source of energy injection in the outer regions of J1652. A significant distinction between the fields of XID2028 and J1652 is the presence of several companion galaxies within 10--15 kpc of the quasar in J1652, engaged in tidal interactions with the quasar host \citep{Wylezalek2022, Vayner2023}, whereas no such companions have been identified near XID2028 with current data. These interactions are a plausible source of the additional energy in J1652. If we assume the quasar is able to impart approximately the same amount of kinetic energy up to a distance of $\sim 10$ kpc based on the results of XID2028, then ratio of $\langle\epsilon\rangle_{\rm outer}/\langle\epsilon\rangle_{\rm inner}$ in J1652 suggests that tidal interactions could inject approximately twice the energy into the gas motions at these distances. 

In addition, this excess of turbulent energy in the outer region of J1652 systemic gas is around an order of magnitude greater than that observed in the J1652 tidal component.  If the measurement of J1652 tidal component reflects the typical energy state post-tidal interactions, the VSF measurements with a higher amplitude from the systemic gas in the outer regions imply that multiple tidal interactions or interactions between outflows and the tidally stripped gas could be contributing to this energy surplus in the host ISM. An alternative possibility is that J1652 recently underwent a major merger, releasing substantial energy into the system, akin to the quasar-galaxy merger recently reported by \cite{Decarli2024}. Based on previous {\it HST} data where a large, extended tidal feature was identified, \cite{Zakamska2019} also proposed J1652 as a likely candidate for a major merger.  


\subsection{Turbulent heating rate and energy cascade timescale}
\begin{figure}
    \includegraphics[width=\linewidth]{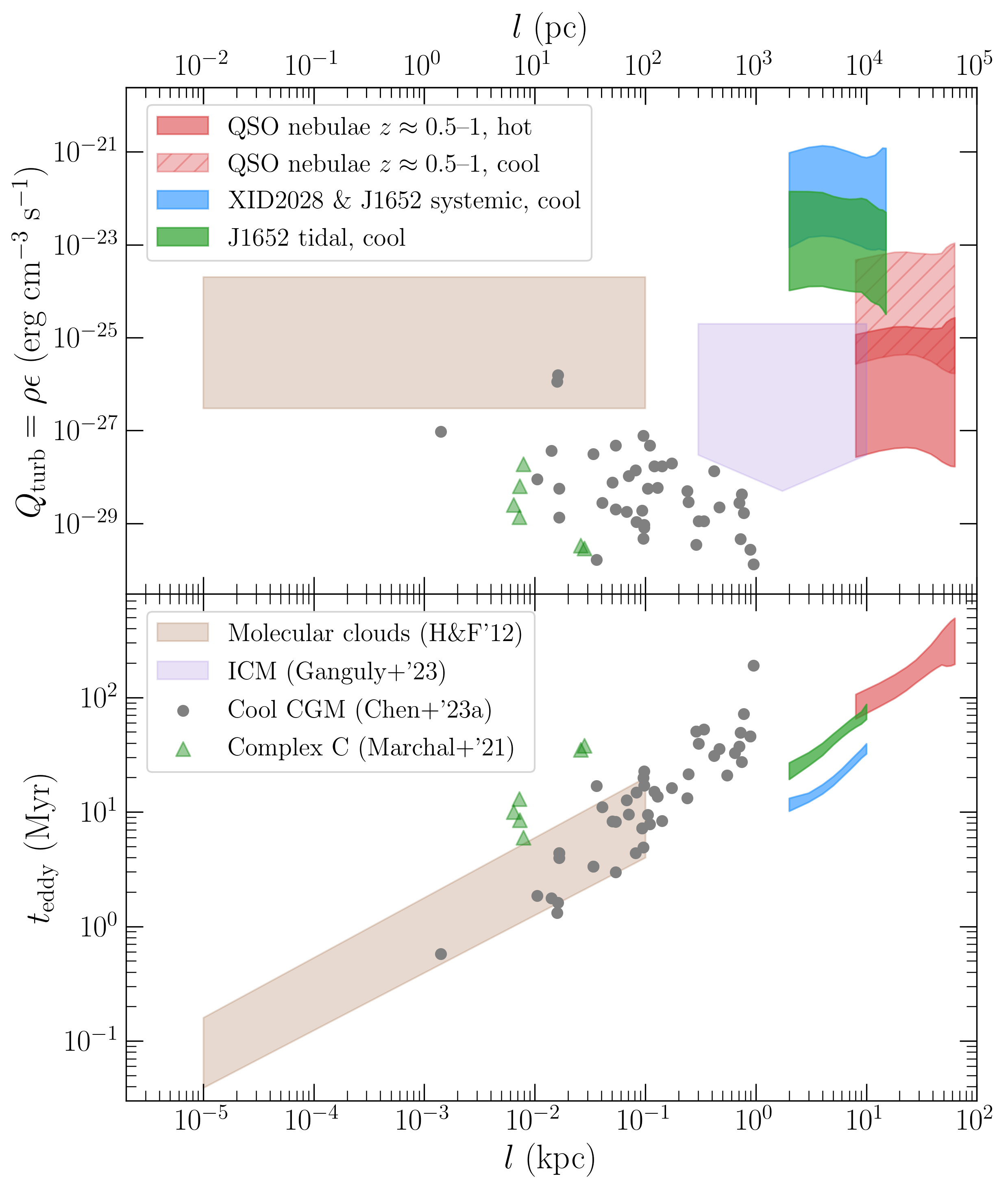}
    \caption{{\it Top panel:} turbulent heating rate per volume, $Q_{\rm turb}$, across various scales and systems. The blue-shaded region represents $Q_{\rm turb}$ for the cool phase gas in XID2028 and the systemic component of J1652 with a density range of 10--500 cm$^{-3}$. The green-shaded region shows the $Q_{\rm turb}$ measured for the tidal gas of J1652. The red and red-hatched regions indicate the estimated $Q_{\rm turb}$ for quasar nebulae at $z\approx 0.5$--1, as presented in \cite{Chen2023, Chen2024}, assuming gas densities of 0.01--1 cm$^{-3}$ for the hot phase and 1--40 cm$^{-3}$ for the cool phase. The gray-shaded region, showing measurements from \cite{Ganguly2023} for the ICM, represents an upper limit due to potential contributions from bulk motions in the utilized velocity maps. The brown-shaded region shows results for star-forming molecular clouds as presented in \cite{Hennebelle2012}.
    \protect\cite{Marchal2021}'s measurements of $Q_{\rm turb}$ for a bright concentration location in the HVC Complex C are shown by the green data points.
    Gray points display the results for CGM cool clumps based on absorption line measurements presented in \citet{Chen:2023b}. The fields XID2028 and J1652 exhibit the highest $Q_{\rm turb}$ among the systems presented here.  {\it Bottom panel:} the eddy turnover time, $t_{\rm eddy}$, for different systems. 
    }
    \label{fig:qturb+teddy}
\end{figure}

While $\epsilon$ reflects the turbulent energy transfer rate per unit mass as discussed above, we can gain further insights into the energetics in these nebulae by multiplying $\epsilon$ with the expected gas density to obtain the turbulent heating rate per unit volume, $Q_{\rm turb}=\rho\,\epsilon$. Based on the [\ion{S}{2}]$\lambda\lambda6717,6731$ doublet line ratio, the median electron density of the nebulae in XID2028 and J1652 is estimated to be $\sim500$ cm$^{-3}$ \citep[][]{Vayner2023, Veilleux2023}. However, this estimate predominantly reflects spaxels with high signal-to-noise detection of [\ion{S}{2}], biasing towards regions of higher density.  To determine a more representative range of electron densities across the nebulae, we consider a fiducial H$\alpha$ surface brightness of $\approx 10^{-16}\,\sbunit$. In most parts of the nebulae, H$\alpha$ emissions are either comparable to or slightly fainter than the [\ion{O}{3}] line, making this surface brightness a suitable benchmark (see Figure 1).  Assuming a temperature of $T\sim 10^4$ K, a recombination coefficient of $\sim 10^{-13}$ cm$^3$ s$^{-1}$ \citep[e.g.][]{Draine2011}, and a cloud depth of $\sim 1$ kpc, we estimate the gas density to be $\sim 10$ cm$^{-3}$, which could increase with a higher clumping factor.
Therefore, we adopt a broad range of 10--500 cm$^{-3}$ to capture the large variability of the gas density. Using this density range, we estimate $Q_{\rm turb}\approx 10^{-23}$--$10^{-21}$ erg cm$^{-3}$ s$^{-1}$ for XID2028 outflow and for J1652 systemic gas, while for J1652 tidal component $Q_{\rm turb}\approx 10^{-24}$--$10^{-22}$ erg cm$^{-3}$ s$^{-1}$, an order of magnitude lower due to the lower power seen in the VSF.
These values of $Q_{\rm turb}$ for both XID2028 outflows and J1652 systemic gas are the highest observed, surpassing previous measurements from a diverse range of physical environments presented in Figure \ref{fig:qturb+teddy}, including star-forming molecular clouds \citep{Hennebelle2012}, a high-velocity cloud (HVC) Complex C in the Milky Way \citep{Marchal2021}, cool CGM clumps probed in absorption \citep{Chen:2023b}, and the ICM from nearby cool-core clusters \citep{Ganguly2023}.  We present the ICM $Q_{\rm turb}$ values from \cite{Ganguly2023} as upper limits in Figure \ref{fig:qturb+teddy} due to the possible presence of strong bulk flows in their data.  The large $Q_{\rm turb}$ values from XID2028 and J1652 reflect the intense turbulent motions of gas in the immediate vicinities of the most luminous quasars at a cosmic epoch of peak AGN feedback. 

We also estimate the eddy turnover time ($t_{\rm eddy}=l/|\Delta v|\approx \langle\Delta v^2\rangle/\epsilon$) and present the values for different systems at the bottom panel of Figure \ref{fig:qturb+teddy}.  $t_{\rm eddy}$ offers insights into the temporal scales of energy dissipation/transfer in turbulent flows. In general, motions at smaller physical scales exhibit shorter timescales, while a wide range of scatter is present for a fixed scale given the wide range of velocities observed in different environments. The tidally stripped gas in the field of J1652 has a $t_{\rm eddy}$ of $\approx 20$--100 Myr, whereas XID2028 outflow and J1652 systemic gas have a relatively short $t_{\rm eddy}$ of $\approx 10$--40 Myr, suggesting more rapid energy transfer processes.  

In addition, consistent with previous findings based on $z\approx 0.5$--1 quasars, our measurements for XID2028 and J1652 suggest that the turbulent energy is subdominant in quasar halos.  Assuming a halo mass of $M_{\rm halo}\approx 10^{13}$--$10^{14}M_\odot$, an NFW profile with a concentration factor of 4, and a baryon fraction of $f_{\rm b}\approx 0.15$ \citep{Planck2020}, the total baryon mass within a radius of 20 kpc from the halo center (i.e., approximate extent of the detected nebulae in this work) is $M$($<$20 kpc)$\sim 10^{10}$--$10^{11}M_\odot$. Adopting $\epsilon\sim 1$ cm$^2$ s$^{-3}$, we obtain a turbulent energy rate of $\dot{E}_{\rm turb}$($<$20 kpc)$=\epsilon\times M$($<20$ kpc)\, $\sim 5\times 10^{44}$ erg s$^{-1}$, which is $\sim$0.1\% of the quasar bolometric luminosity. This ratio is similar to the wind energy fraction estimated for J1652 \citep{Vayner2024} as well as other AGN outflows \citep[e.g.][]{Fabian2012,Sun2017}. 

\section{Discussion and Conclusions}
\label{sec:discussion_conclusion}
In this {\it Letter}, we studied the turbulence and energy injection in the extended nebulae surrounding two luminous quasars, XID2028 ($z=1.5933$) and J1652 ($z=2.9489$), using \textit{JWST}/NIRSpec IFU observations. We leveraged the refined spatial resolution in the near-IR offered by \textit{JWST} to map the velocity fields of the ionized gas around these quasars, tracing line-emitting gas up to $\approx 20$ kpc from the central engine (see Figure \ref{fig:nb_and_velocity}). Our analysis focused on the second-order VSFs to quantify the turbulent motions within these environments, providing insights into the underlying drivers of the gas dynamics. For the first time, we provide empirical constraints on the scale and amplitude of kinetic energy injection from AGN outflows to the surrounding gaseous environment and present direct evidence of outflows interacting with tidal features. 

After accounting for the contribution from the large-scale coherent velocity gradient in the plane of the sky, we observed that the VSFs for both XID2028 and J1652 exhibit a notable flat trend across the scale from $\approx 3$ kpc to $\approx 10$--20 kpc (see Figure \ref{fig:s2}).  A plausible explanation for the flattening in the VSFs is that the energy injection in these nebulae occurs at a scale of $\lesssim$ 10 kpc.  This scenario correlates closely with observations such as the jet-driven bubble with an approximate size of 5 kpc in XID2028 \citep{Cresci2023} and the fast outflows in J1652, which is confined within about 5 kpc from the quasar \citep{Vayner2023,Vayner2024}. Both phenomena are likely major contributors to the observed turbulence and are capable of transferring kinetic energy to the surrounding gas on scales of a few kpc. 

For J1652, the VSFs measured from the systemic gas have an amplitude $\approx 10\times$ higher than that obtained from the tidally-stripped gas (see Figure \ref{fig:s2}), suggesting that AGN activities can significantly boost the turbulent motions in the ISM of the host galaxy and supporting the widely accepted theory that AGN feedback plays a pivotal role in shaping gas properties in galaxies. In addition, the observed differences in VSF amplitudes between the inner and outer regions of J1652 systemic gas hint at additional energy injection mechanisms in the outer regions (see Figure \ref{fig:s2_inner_vs_outer}). The interactions between outflows and nearby companion galaxies, as well as potential past major mergers, are likely contributors to this enhanced turbulence, underscoring the complex interplay between galaxy interactions and AGN feedback.

When compared to their lower-redshift counterparts at $z\approx 0.5$--1, the high-redshift obscured quasars XID2028 and J1652 show significantly higher VSF amplitudes, especially in the systemic and outflow components (see Figure \ref{fig:s2}). The estimated energy transfer rates and turbulent heating rates are exceptionally high in these environments (see Table \ref{tab:epsilon_summary} and Figure \ref{fig:qturb+teddy}), highlighting the energetic motions of gas near powerful AGNs and echoing the discussion above. 

Future observations are crucial to unraveling the physical origins of the observed differences between the obscured quasars and the $z\approx 0.5$--1, UV-bright quasars. These observations need to target UV-bright quasars and resolve the inner $\sim 10$ kpc region, not only consolidating the detection of the characteristic energy injection scale of $\lesssim 10$ kpc but also illuminating whether UV-bright quasars exhibit different VSF properties close to the central engine, compared to the obscured quasars studied here. These potential variations might highlight population differences between UV-bright and obscured quasars, the latter of which may represent an early, heavily obscured phase of quasar activity, offering insights into the diverse evolutionary stages of quasars. 
Additionally, the role of cosmic evolution in ISM/CGM turbulence, particularly at higher redshifts where AGN feedback is more intense, warrants further investigation. Expanding our sample to include more high-redshift quasars will be essential in addressing these questions.  

In conclusion, our study highlights the complex interplay between AGN activities and the kinematics of the surrounding gas.  These early-stage findings offer critical insights that help set the foundation for forthcoming larger-scale analyses. The detailed observations into the turbulent dynamics offered by the unprecedented spatial resolution of \textit{JWST}/NIRSpec IFU data at high redshifts provide a valuable framework for further theoretical and observational efforts aimed at unraveling the processes that govern galaxy evolution in the presence of powerful AGN feedback. Future investigations will benefit significantly from integrating multi-scale and multi-epoch data to build a more comprehensive picture of how AGNs influence their host galaxies and beyond.


\begin{acknowledgments}
The authors thank Sean Johnson and Irina Zhuravleva for their helpful comments on this work. HWC and MCC acknowledge partial support from The University of Chicago Women’s Board grants and from The University of Chicago Data Science Institute (DSI). MCC is supported by the Brinson Foundation through the Brinson Prize Fellowship Program. NLZ is supported in part by NASA through STScI grant JWST-ERS-01335. The data used in this work are publicly available on the Mikulski Archive for Space Telescopes (MAST). The NIRSpec IFU data for SDSSJ1652 can be accessed via doi:\dataset[10.17909/qacq-9285]{http://dx.doi.org/10.17909/qacq-9285}, while the data for XID2028 is at doi:\dataset[10.17909/04tb-mn90]{http://dx.doi.org/10.17909/04tb-mn90}.
This research has made use of the services of the ESO Science Archive Facility and the Astrophysics Data Service (ADS)\footnote{\url{https://ui.adsabs.harvard.edu/classic-form}}. The analysis in this work was greatly facilitated by the following \texttt{python} packages:  \texttt{Numpy} \citep{Numpy}, \texttt{Scipy} \citep{Scipy}, \texttt{Astropy} \citep{astropy:2013,Astropy2018, Astropy2022}, \texttt{Matplotlib} \citep{Matplotlib}, and \texttt{MPDAF} \citep{MPDAF}.  This work was completed with resources provided by the University of Chicago Research Computing Center.
\end{acknowledgments}

\clearpage

\bibliography{main}{}
\bibliographystyle{aasjournal}


\end{document}